\documentclass[%
 reprint,
 amsmath,amssymb,amsthm,
 nofootinbib,
 aps,
]{revtex4-1}

\usepackage{dcolumn}% Align table columns on decimal point
\usepackage{bm}% bold math
\usepackage[mathlines]{lineno}
\usepackage{graphicx}
\usepackage{csquotes}
\usepackage[final]{changes}

\newtheorem{Theorem}{Theorem}

\newtheorem{lemma}{Lemma}
	
\begin{document}
\preprint{APS/123-QED}

%\title{How is Entanglement Preserved and Manifested Between Two Spatially Separated Particles?}
\title{On the Preservation and Manifestation of Quantum Entanglement}

\author{Jianhao M. Yang}
\email[]{jianhao.yang@alumni.utoronto.ca}
\affiliation{Qualcomm, San Diego, CA 92321, USA}

\date{\today}		

\begin{abstract}
Bell experiments have confirmed that quantum entanglement is an inseparable correlation but there is no faster-than-light influence between two entangled particles when a local measurement is performed. However, how such an inseparable correlation is maintained and manifested when the two entangled particle are space-like separated is still not well understood. The recently proposed extended least action principle for quantum mechanics brings new insights to this question. By applying this principle, we show here that even though the inseparable correlation may be initially created by previous physical interaction between the two particles, the preservation and manifestation of such inseparable correlation are achieved through extremizing an information metric that measures the additional observable information of the bipartite system due to vacuum fluctuations. This is physically realized even though there is no further interaction when the two particles move apart, and the underlying vacuum fluctuations are local. In other words, the propagation of inseparable correlation in quantum theory is realized by an information requirement and through a local mechanism. An example of two entangled free particles described by Gaussian wave packets is provided to illustrate these results. 
\end{abstract}
%In other words, Bell inseparability is an informational consequence due to local vacuum fluctuations. 
\maketitle
%\onecolumngrid
%========================================
%========================================
\section{Introduction}
Entanglement is one of the most intrinsic features of quantum mechanics. It is used extensively as an unique resource for enabling quantum computing and quantum information processing~\cite{Nielsen,Hayashi15}. However, there is still mystery on the underlying mechanism of entanglement as it appears demonstrating the existence of non-local causal relation between two space-like separated entangled particles. To clearly show this conceptual difficulty, let's consider the Bohm's version of the EPR thought experiment~\cite{EPR}. Suppose a pair of electrons $A$ and $B$ were previously interacting so that their spins are maximally entangled but now are spatially separated. Observer Alice is collocated with $A$, while observer Bob is with $B$. When Alice measures $A$ with outcome of spin up along $z$-axis, Bob measures $B$ along the $z$-axis and also gets the outcome of spin up. Now Alice measures $A$ along a different $z'$-axis tilted with an angle from $z$-axis, with outcome of spin down. Meanwhile, Bob measures $B$ along the $z'$-axis as well and also get the outcome of spin down. Alice's measurement outcome on $A$ is completely random, say, she will get get spin up half chance and spin down half chance. With the information of Alice's measurement outcome on $A$, Bob always obtains the same measurement outcome on $B$. Several fundamental questions are in order. 1.) The measurement outcome of $B$ appears depending on the measurement outcome of the remotely separated $A$, why? Is there a ``spooky action at a distance" that cause $B$ to change the spin property? 2.) Furthermore, since the measurement outcome of $A$ is completely random, then exactly what is the value of spin for $B$ before Alice's measurement on $A$? 3.) Why and how such inseparable correlation between $A$ and $B$ is preserved and manifested even though they are remotely separated? 

Parts of these questions have been answered and resolved in current quantum theory. The first and second questions can be partially resolved if one admits that a physical property is only valid when measurement on the observable is performed. There is no definite physical property prior to and independent of observation. This indeed aligns with the Copenhagen interpretation. Furthermore, Bell inequality and its experimental confirmation~\cite{Bell, Hensen} reveal that quantum theory cannot be both local (physical influences do not propagate faster than light) and realistic (physical properties are defined prior to and independent of observation). Violation of Bell inequality in quantum entanglement just shows that the correlation between two remotely separated entangled systems is inseparable. However, it does not necessarily imply there is faster-than-light influence of one system on the other. Instead, these correlations may simply reveal some dependence relation between the two systems which was established when they interacted in the past. This kind of correlation is referred to as Bell nonlocality~\cite{Brunner}, or, more accurately, Bell non-separability~\cite{Hall2015}. But there is still an open question - why and how such non-separable correlation is preserved and manifested even though the two subsystems are remotely separated and there is no direct influence between them? This is the third question mentioned earlier. Mathematically, one can solve the Schr\"{o}dinger equation for a bipartite system with appropriate initial conditions to obtain an entangled quantum state. But conceptually this is not satisfactory because it does not provide physical insights when answering the question.

The purpose of this paper is to investigate the answers to the third question. The investigation is based on the extended least action principle for quantum mechanics~\cite{Yang2023}. 
The principle extends the least action principle in classical mechanics by factoring two assumptions. First, there is a lower limit to the amount of action a physical system needs to exhibit in order to be observable. Such a discrete action unit is defined by the Planck constant. It serves as a basic unit in terms of action to measure the observable information a physical system exhibits during its motion. Second, there is vacuum fluctuation that is completely random. New information metric, $I_f$, is introduced to measure the additional observable information due to these random fluctuations, which is then converted to additional amount of action due to the vacuum fluctuations using the first assumption. The laws of dynamics can then be obtained by extremizing the total actions from both classical trajectory and the vacuum fluctuations. By recursively applying the principle in an infinitesimal time interval and an accumulated time interval, the uncertainty relation and the Schr\"{o}dinger equation are recovered respectively. The crucial component of this formulation is that the quantumness can be attributed to the new information metrics $I_f$. %$I_f$ is defined as the information distance between probability distribution due to vacuum fluctuations and probability distribution without vacuum fluctuations, and importantly, such vacuum fluctuations are local. 

By applying the extended least action principle to a bipartite quantum system, we show that the information metrics $I_f$ play a critical role in quantum entanglement. Although the initial inseparable correlation in a bipartite system is created by previous physical interaction, the preservation and manifestation of such inseparable correlation are realized without further interaction even if the two subsystems are moving apart. This is due to the constraint to extremize $I_f$ in addition to extremizing the classical action. But $I_f$ is just an information metric introduced to measure the observable information due to the vacuum fluctuations of $A$ and $B$, and these underlying vacuum fluctuations are local and independent from each other. Thus, we confirm that even though entanglement is an inseparable correlation, the preservation and manifestation aspects of entanglement are not necessarily causal. To make this point more comprehensible, an example of two entangled free particles described by Gaussian wave packets is analyzed in Section \ref{sec:Gaussian}.

The extended least action principle was inspired by the recent interests in searching for foundational principles of quantum theory from the information perspective~\cite{Rovelli:1995fv, zeilinger1999foundational, Brukner:ys, Brukner:1999qf, Brukner:2002kx, Fuchs2002, brukner2009information, Brukner:vn, spekkens2007evidence, Spekkens:2014fk, Paterek:2010fk, gornitz2003introduction, lyre1995quantum, Hardy:2001jk, masanes2011derivation, Mueller:2012ai, Masanes:2012uq, chiribella2011informational, Mueller:2012pc, Hardy:2013fk, kochen2013reconstruction, 2008arXiv0805.2770G, Hall2013, Hoehn:2014uua, Hoehn:2015, Stuckey, Mehrafarin2005, Caticha2011, Caticha2019, Frieden, Reginatto}. One of the motivations is to reformulate quantum mechanics with information theoretic principles such that some of the conceptual challenges can be resolved. Applying the principle here to gain new insights on quantum entanglement helps us to answer the third question both conceptually and mathematically.

The rest of the article is organized as follows. In Section \ref{LIP}, we review the extended least action principle, the underlying assumptions, and how the basic quantum theory is derived from it. In Section \ref{sec:entBipartite}, the principle is applied to derive the Schr\"{o}dinger equation for a non-interacting bipartite system. It is shown in Theorem 1 that two conditions, inseparable initial joint probability distribution and $I_f\ne 0$, are necessary for preserving and manifesting entanglement. This is followed by a detailed breakdown on the processes of quantum entanglement. Section \ref{sec:examples} provides two examples on Theorem 1. In particular, the joint probability distribution of two entangled free particles represented by Gaussian wave packets is computed and visually shown. Section \ref{sec:discussion} discusses the implications of $I_f$ on locality and causality, and how the EPR paradox can be resolved.

\section{Extending the Leas Action Principle}
\label{LIP}
\subsection{The Principle and Its Underlying Assumptions}
Ref.~\cite{Yang2023} shows that the least action principle in classical mechanics can be extended to derive quantum formulation by factoring in the following two assumptions.
\begin{displayquote}
\emph{Assumption 1 -- A quantum system experiences vacuum fluctuations constantly. The fluctuations are local and completely random.}
\end{displayquote}
\begin{displayquote}
\emph{Assumption 2 -- There is a lower limit to the amount of action that a physical system needs to exhibit in order to be observable. This basic discrete unit of action effort is given by $\hbar/2$ where $\hbar$ is the Planck constant.}
\end{displayquote}

The first assumption is generally accepted in mainstream quantum mechanics, which is responsible for the intrinsic randomness of the dynamics of a quantum object. Even though we do not know the physical details of the vacuum fluctuation, the crucial assumption here is the locality of the vacuum fluctuation. This implies that for a composite system, the fluctuation of each subsystem is independent of each other. The precise definition of locality is given in Section \ref{sec:entBipartite}. 

The implications of the second assumption need more elaborations. Historically the Planck constant was first introduced to show that energy of radiation from a black body is discrete. One can consider the discrete energy unit as the smallest unit to be distinguished, or detected, in the black body radiation phenomenon. In general, it is understood that Planck constant is associated with the discreteness of certain observables in quantum mechanics. Here, we instead interpret the Planck constant from an information measure point of view. Assumption 2 states that there is a lower limit to the amount of action that the physical system needs to exhibit in order to be observable or distinguishable in potential observation, and such a unit of action is determined by the Planck constant. 

Making use of this understanding of the Planck constant conversely provides us a new way to calculate the additional action due to vacuum fluctuations. That is, even though we do not know the physical details of vacuum fluctuations, the vacuum fluctuations manifest themselves via a discrete action unit determined by the Planck constant as an observable information unit. If we are able to define an information metric that quantifies the amount of observable information manifested by vacuum fluctuations, we can then multiply the metric with the Planck constant to obtain the action associated with vacuum fluctuations. Then, the challenge to calculate the additional action due to vacuum fluctuation is converted to define a proper new information metric $I_f$, which measures the additional distinguishable, hence observable, information exhibited due to vacuum fluctuations. Even though we do not know the physical details of vacuum fluctuations (except that as Assumption 1 states, these vacuum fluctuations are completely random and local), the problem becomes less challenged since there are information-theoretic tools available. The first step is to assign a transition probability distribution due to vacuum fluctuation for an infinitesimal time step at each position along the classical trajectory. The distinguishability of vacuum fluctuation then can be defined as the information distance between the transition probability distribution and a uniform probability distribution. Uniform probability distribution is chosen here as reference to reflect the complete randomness of vacuum fluctuations. In information theory, the common information metric to measure the information distance between two probability distributions is relative entropy. Relative entropy is more fundamental to Shannon entropy since the latter is just a special case of relative entropy when the reference probability distribution is a uniform distribution. But there is a more important reason to use relative entropy. As shown in later sections, when we consider the dynamics of the system for an accumulated time period, we assume the initial position is unknown but is given by a probability distribution. This probability distribution can be defined along the position of classical trajectory without vacuum fluctuations, or with vacuum fluctuations. The information distance between the two probability distributions gives the additional distinguishability due to vacuum fluctuations. It is again measured by a relative entropy. Thus, relative entropy is a powerful tool allowing us to extract meaningful information about the dynamic effects of vacuum fluctuations. Concrete form of $I_f$ will be defined later as a functional of Kullback-Leibler divergence $D_{KL}$, $I_f:=f(D_{KL})$, where $D_{KL}$ measures the information distances of different probability distributions caused by vacuum fluctuations. Thus, the total action from classical path and vacuum fluctuation is
\begin{equation}
\label{totalAction}
    S_t = S_c + \frac{\hbar}{2}I_f,
\end{equation}
where $S_c$ is the classical action. Non-relativistic quantum theory can be derived through a variation approach to minimize such a functional quantity, $\delta S_t=0$. When $\hbar \to 0$, $S_t=S_c$. Minimizing $S_t$ is then equivalent to minimizing $S_c$, resulting in Newton's laws in classical mechanics. However, in quantum mechanics, $\hbar \ne 0$, the contribution from $I_f$ must be included when minimizing the total action. We can see $I_f$ is where the quantum behavior of a system comes from. These ideas can be condensed as
\begin{displayquote}
\emph{\textbf{Extended Least Action Principle} -- The law of physical dynamics for a quantum system tends to exhibit as little as possible the action functional defined in (\ref{totalAction}).}
\end{displayquote}

%The existence of the Planck constant as a unit of action for the physical system to exhibit in order to be observable allows us to ask the question: Given a certain amount of action $S_c$ from its motion along a classical trajectory, how much observability does the system exhibit from its dynamics? According to assumption 2, this is calculated as $I_p = 2S_c/\hbar$. $I_p$ is not a conventional information metric but it has clear meaning about a piece of physical information. That is, it can be considered as the amount of observable information measured in the unit of $\hbar/2$. This step of converting $S_c$ into $I_p$ appears trivial mathematically, but conceptually it is not. It recasts the least action principle into a least observability principle, and shifts the working language to be information related. Thus, $I_p$ can be paired with additional information metrics due to vacuum fluctuations. To measure the degree of observability due to vacuum fluctuations, a new information metric $I_f$ is introduced. $I_f$ is defined as a metric to measure the additional distinguishable, hence observable, information exhibited due to vacuum fluctuations. More specifically, $I_f$ is defined as a functional of Kullback-Leibler divergence $D_{KL}$, $I_f:=f(D_{KL})$, where $D_{KL}$ measures the information distances of different probability distributions caused by vacuum fluctuations. Thus, the total degree of observability due to both classical trajectory and vacuum fluctuation is

Alternatively, we can interpret the extended least action principle more from an information perspective by rewriting (\ref{totalAction}) as 
\begin{equation}
\label{totalInfo}
    I_t =\frac{2}{\hbar} S_c + I_f,
\end{equation}
where $I_t=2S_t/\hbar$. Denote $I_p=2S_c/\hbar$, which measures the amount of $S_c$ using the discrete unit $\hbar/2$. $I_p$ is not a conventional information metric but can be considered carrying meaningful physical information about the observability of the classical trajectory. To see this connection, recall that the classical action is defined as an integral of the Lagrangian over a period of time along a path trajectory of a classical object. There are two aspects to understanding the action functional. In classical mechanics, the path trajectory can be traced, measured, or observed. Given two fixed end points, the longer the path trajectory, the larger the value of the action. It indicates 1.) the more dynamic effort the system exhibits; and 2.) the easier to trace the path and distinguish the object from the background reference frame, or in other words, the more physical information available for potential observation. Thus, action $S_c$ not only quantifies the dynamic effort of the system, but also is associated with the detectability, or observability, of the physical object during the dynamics along the path. In classical mechanics, we focus on the first aspect via the least action principle, and derive the law of dynamics from minimizing the action effort. The second aspect is not useful in practice since we cannot quantify the intuition that $S$ is associated with the observability of the physical object. One reason is that there is no natural unit of action to convert $S$ into an information related metric. The introduction of the Planck constant in Assumption 2 helps to quantify this intuition. Similarly, $I_f$ measures the distinguishable information of the probability distributions with and without field fluctuations. Thus, $I_t$ is the total observable information. With (\ref{totalInfo}), the extended least action principle can be restated as\footnote{The term observability is not related to the concept of observable in traditional quantum physics since it is not associated with a Hermitian operator. Also, one should not confuse the term with the same terminology in system control theory.} 
\begin{displayquote}
\emph{\textbf{Principle of Least Observability} -- The law of physical dynamics for a quantum system tends to exhibit as little as possible the observable information defined in (\ref{totalInfo}).}
\end{displayquote}

Mathematically, there is no difference between (\ref{totalAction}) and (\ref{totalInfo}) when applying the variation principle to derive the laws of dynamics. The form of (\ref{totalAction}) in terms of actions appears more familiar in the physics community. However, The form of (\ref{totalInfo}) in terms of observability seems conceptually more generic. We will leave the exact interpretations of the principle alone and use the two interpretations interchangeable in this paper. The key point to remember is that the Planck constant connects the physical action to metrics related to observable information in either interpretation.

%Non-relativistic quantum formulation can be derived through a variation method that demands $I$ is stationary, that is, $\delta I=0$. When $\hbar \to 0$, the observability due to classical path $I_p \to \infty$. Thus, the system can be observed with infinite accuracy, and any finite amount of $I_f$ can be ignored. Minimizing $I$ is then equivalent to minimizing $S_c$, resulting in the dynamics laws of classical mechanics. However, in quantum mechanics, according to the Assumption 2, the action to exhibit observability is discrete so that $\hbar \ne 0$, and $I_p$ is finite. This means there is only a finite amount of observable information available. The contribution from $I_f$ can be comparable to $I_f$ and therefore must be included when minimizing the total amount of observable information. These ideas can be summarized as\footnote{The principle can be called the principle of least distinguishability as well, since the term distinguishability and observability is interchangeable in this paper. Also, the term observability should not be confused with the same terminology in system control theory.}

\subsection{Basic Quantum Formulation}
\label{sec:shorttime}
Basic quantum formulation can be derived by recursively applying the least observability principle in two steps. Without loss of generality, only one dimensional systems are considered. First, we consider the dynamics of a system with an infinitesimal time internal $\Delta t$ due to vacuum fluctuation only. Define the probability for the system to transition from a space position ${x}$ to another position ${x}+{w}$, where ${w}=\Delta {x}$ is the displacement due to fluctuations, as $\wp({x}+{w}|{x})d{w}$. The expectation value of classical action is $S_c=\int \wp({x}+{w}|{x})Ld{w}dt$. Since we only consider the vacuum fluctuations, the Lagrangian $L$ only contains the kinetic energy, $L=\frac{1}{2}mv^2$. For an infinitesimal time internal $\Delta t$, one can approximate the velocity ${v}={w}/\Delta t$. This gives 
\begin{equation}
\label{action1}
    S_c=\frac{m}{2\Delta t}\int^{+\infty}_{-\infty} \wp({x}+{w}|{x})w^2 d{w}.
\end{equation}
The information metrics $I_f$ is defined as the Kullback–Leibler divergence, to measure the information distance between $\wp({x}+{w}|{x})$ and a uniform prior probability distribution $\mu$ that reflects the vacuum fluctuations are completely random with maximal ignorance~\cite{Caticha2019, Jaynes}, 
\begin{align*}
    I_f  &=: D_{KL}(\wp({x}+{w}|{x}) || \mu) \\
    &= \int \wp({x}+{w}|{x})ln[\wp({x}+{w}|{x})/\mu]d{w}.
\end{align*}
Insert both $S_c$ and $I_f$ into (\ref{totalInfo}) and perform the variation procedure, one obtain
\begin{equation}
\label{transP}
    \wp({x}+{w}|{x}) = \frac{1}{Z}e^{-\frac{m}{\hbar\Delta t}w^2},
\end{equation}
where $Z$ is a normalization factor. Equation (\ref{transP}) shows that the transition probability density is a Gaussian distribution. The variance $\langle w^2\rangle = \hbar\Delta t/2m$. Recalling that $w/\Delta t = v$ is the approximation of velocity due to the vacuum fluctuations, one can deduce
\begin{equation}
\label{exactUR}
    \langle\Delta x\Delta p\rangle = \frac{\hbar}{2}.
\end{equation}
Applying the Cauchy–Schwarz inequality gives
\begin{equation}
    \langle\Delta x\rangle\langle\Delta p\rangle \ge \hbar/2.
\end{equation}

In the second step, we will derive the dynamics for a cumulative period from $t_A\to t_B$. In classical mechanics, the equation of motion is described by the Hamilton-Jacobi equation, 
\begin{equation}
    \label{HJE}
    \frac{\partial S}{\partial t }+ \frac{1}{2m}\nabla S\cdot\nabla S + V = 0.
\end{equation}
Suppose the initial condition is unknown, and define $\rho ({x}, t)$ as the probability density for finding a particle in a given volume of the configuration space. The probability density must satisfy the normalization condition $\int \rho ({x}, t) d{x} = 1$, and the continuity equation 
\begin{equation*}
    \frac{\partial\rho ({x}, t)}{\partial t }+ \frac{1}{m}\nabla \cdot(\rho ({x}, t)\nabla S) = 0.
\end{equation*}
The pair $(S, \rho)$ completely determines the motion of the classical ensemble. As pointed out by Hall and Reginatto~\cite{Hall:2001,Hall:2002}, the Hamilton-Jacobi equation, and the continuity equation, can be derived from classical action
\begin{equation}
    \label{cAction}
    S_c = \int\rho\{ \frac{\partial S}{\partial t} + \frac{1}{2m}\nabla S\cdot\nabla S + V\} d{x}dt
\end{equation}
through fixed point variation with respect to $\rho$ and $S$, respectively. Note that $S_c$ and $S$ are different physical variables, where $S_c$ can be considered as the ensemble average of classical action and $S$ is a generation function that satisfied $\mathbf{p}=\nabla S$~\cite{Yang2023}. 

To define the information metrics for the vacuum fluctuations, $I_f$, we slice the time duration $t_A\to t_B$ into $N$ short time steps $t_0=t_A, \ldots, t_j, \ldots, t_{N-1}=t_B$, and each step is an infinitesimal period $\Delta t$. In an infinitesimal time period at time $t_j$, the particle not only moves according to the Hamilton-Jacobi equation but also experiences random fluctuations. Such additional revelation of distinguishability due to the vacuum fluctuations on top of the classical trajectory is measured by the following definition,
\begin{align}
\label{DLDivergence}
    I_f &=: \sum_{j=0}^{N-1}\langle D_{KL}(\rho ({x}, t_j) || \rho ({x}+{w}, t_j))\rangle_w \\
    &=\sum_{j=0}^{N-1}\int d{w}\wp({w})\int dx\rho ({x}, t_j)ln \frac{\rho ({x}, t_j)}{\rho ({x}+{w}, t_j)}.
\end{align}
When $\Delta t\to 0$, $I_f$ turns out to be~\cite{Yang2023}
\begin{equation}
\label{FisherInfo}
    I_f = \int d{x}dt \frac{\hbar}{4m}\frac{1}{\rho}\nabla\rho \cdot \nabla\rho.
\end{equation}
Eq. (\ref{FisherInfo}) contains the term related to Fisher information for the probability density~\cite{FriedenBook} but bears much more physical significance than Fisher information. %It shows that $I_f$ is proportional to $\hbar$. This is not trivial because it avoids introducing additional arbitrary constants for the subsequent derivation of Schr\"{o}dinger equation. Defining $I_f$ using other generic forms of relative entropy such as R\'{e}nyi divergence or Tsallis divergence, one will obtain a generalized Schr\"{o}dinger equation. 
Inserting (\ref{cAction}) and (\ref{FisherInfo}) into (\ref{totalInfo}), and performing the variation procedure on $I$ with respect to $S$ gives the continuity equation, while variation with respect to $\rho$ leads to the quantum Hamilton-Jacobi equation,
\begin{equation}
\label{QHJ}
    \frac{\partial S}{\partial t} + \frac{1}{2m}\nabla S\cdot\nabla S + V - \frac{\hbar^2}{2m}\frac{\nabla^2\sqrt{\rho}}{\sqrt{\rho}} = 0,
\end{equation}
Defined a complex function $\Psi=\sqrt{\rho}e^{iS/\hbar}$, the continuity equation and the extended Hamilton-Jacobi equation (\ref{QHJ}) can be combined into a single differential equation,
\begin{equation}
    \label{SE}
    i\hbar\frac{\partial\Psi}{\partial t} = [-\frac{\hbar^2}{2m}\nabla^2 + V]\Psi,
\end{equation}
which is the Schr\"{o}dinger Equation. 

The last term in (\ref{QHJ}) is the Bohm quantum potential~\cite{Bohm1952}. The Bohm potential is considered responsible for the non-locality phenomenon in quantum mechanics~\cite{Bohm2}. Historically, its origin is mysterious. Here we show that it originates from the information metrics related to relative entropy, $I_f$. This is a key finding as it will play a crucial role in explaining the entanglement phenomenon in the next section.

\section{Entanglement of Two Free Particles}
\label{sec:entBipartite}
In this section, the least observability principle is applied to study two entangled but non-interacting particles. We will show that the information metrics $I_f$ plays a pivot role in preserving and manifesting entanglement. Consider two subsystems $A$ and $B$ with masses $m_a$ and $m_b$, respectively. At $t=0$, the interaction between them is turned off so that the external potential $V({x}_a, {x}_b)=V_a({x}_a) + V_b({x}_b)$ and the two subsystems are moving away from each other. But there can be interaction between them for $t < 0$ such that the joint probability density at $t=0$ is inseparable, 
\begin{equation}
    \label{initCond}
    \rho({x}_a, {x}_b, t=0) \ne \rho_a({x}_a, t=0)\rho_b({x}_b, t=0).
\end{equation}
On the other hand, if there is no interaction between them for $t < 0$, the joint probability density at $t=0$ can be separable, 
\begin{equation}
    \label{initCond2}
    \rho({x}_a, {x}_b, t=0) = \rho_a({x}_a, t=0)\rho_b({x}_b, t=0).
\end{equation}
Both initial conditions must be considered.

\subsection{Classical Dynamics}
If $A$ and $B$ are two classical particles, there is no vacuum fluctuation so that $I_f=0$. Since there is no interaction between $A$ and $B$ for $t\ge 0$, the momentum of each subsystem is independent from each other. That is, $p=p_a+p_b$. Recall that $p=\nabla S$, $S$ must be separable so that $S({x}_a, {x}_b, t)=S_a({x}_a, t)+S_b({x}_b, t)$. Consequently, the ensemble average of classical action is
\begin{equation}
    \label{cAction2}
    \begin{split}
    S_c = &\int\rho({x}_a, {x}_b, t)\{ (\frac{\partial S_a}{\partial t} + \frac{1}{2m_a}\nabla_a S_a\cdot\nabla_a S_a + V_a) \\
    & + (\frac{\partial S_b}{\partial t} + \frac{1}{2m_b}\nabla_b S_b\cdot\nabla_b S_b + V_b)\} d{x}_ad{x}_bdt.
    \end{split}
\end{equation}
In the case that the probability density $\rho(x_1, x_2, t)$ is inseparable, the law of marginal probability still holds\footnote{The law of marginal probability always holds in classical probability, but is not necessarily true in quantum theory. The law of marginal probability is true if the two measurement operations are commutative~\cite{Yang2022}. For two subsystems that are space-like separated, measurements on each subsystem are commutative.},
\begin{align}
\label{lmp}
    \int \rho({x}_a, {x}_b, t)d{x}_a &= \rho_b({x}_b, t), \\
    \int \rho({x}_a, {x}_b, t)d{x}_b &= \rho_a({x}_a, t).
\end{align}
The above identities certainly hold in the case that the joint probability density is separable. Thus, regardless the separability of the joint probability density, $S_c = (S_c)_a + (S_c)_b$, with 
\begin{equation}
\label{SC1}
    (S_c)_a = \int\rho_a({x}_a, t) (\frac{\partial S_a}{\partial t} + \frac{1}{2m_a}\nabla_a S_a\cdot\nabla_a S_a + V_a)d{x}_adt
\end{equation}
and a similar expression for $(S_c)_b$. The Hamilton-Jacobi equation and the continuity equation for each subsystem can be derived independently from variations on $(S_c)_a$ and $(S_c)_b$, respectively. The dynamics of the two classical particles are independent. There is no impact on the dynamics of the two particles even if $\rho({x}_a, {x}_b, t) \ne \rho({x}_a, t)\rho({x}_b, t)$.

\subsection{Quantum Dynamics}
Now we apply the least observability principle to a bipartite system. The ensemble average of classical action for the bipartite system is given by (\ref{cAction2}). In addition, we need to consider the information metric $I_f$ for the bipartite system due to fluctuations. One of the key points in Assumption 1 is the locality of the vacuum fluctuations. The fluctuations experienced by particle A are completely independent from the fluctuations experienced by particle B. Formally, the locality of vacuum fluctuation can be defined by the separability of the joint transition probability of the bipartite system,
\begin{equation}
\label{jointTP}
\begin{split}
    \wp({x}_a & +{w}_a, {x}_b+{w}_b, t_j|{x}_a,{x}_b, t_j) = \\
    &\wp_a({x}_a+{w}_a, t_j|{x}_a, t_j)\wp_b({x}_b+{w}_b, t_j|{x}_b, t_j).
\end{split}
\end{equation}
Extend the definition of $I_f$ in (\ref{DLDivergence}) to the bipartite system:
\begin{equation}
\label{DLDivergencefor2}
    I_f =: \sum_{j=0}^{N-1}\langle D_{KL}(\rho (x_a, x_b, t_j) || \rho ({x}_a+{w}_a, {x}_b+{w}_b, t_j)\rangle_w.
\end{equation}
Using (\ref{jointTP}), we show in Appendix A that when $\Delta t \to 0$,
\begin{equation}
\label{If3}
    I_f = \int d{x}_ad{x}_bdt\{\frac{\hbar}{4m_a}\frac{\nabla_a\rho\cdot\nabla_a\rho}{\rho} + \frac{\hbar}{4m_b}\frac{\nabla_b\rho\cdot\nabla_b\rho}{\rho}\}.
\end{equation}
Variation of $I_f$ with respect to $\rho$ gives the Bohm quantum potential for the bipartite system, as shown in (A3) of Appendix A,
\begin{equation}
\label{BohmQ}
    Q = - \frac{\hbar^2}{2m_a}\frac{\nabla_a^2\sqrt{\rho}}{\sqrt{\rho}} - \frac{\hbar^2}{2m_b}\frac{\nabla_b^2\sqrt{\rho}}{\sqrt{\rho}}.
\end{equation}
The interesting finding here is that even though the vacuum fluctuations for the two subsystems are independent from each other, $I_f$ and the Bohm potential are inseparable in general. The inseparability depends on the inseparability of the initial condition $\rho({x}_a, {x}_b, t)$. This suggests that there is no need for a non-local mechanism underlying the inseparability of the Bohm quantum potential\footnote{Although (\ref{BohmQ}) in mathematical form is identical to the Bohm quantum potential for a bipartite system, the underlying physics is very different from the Bohmian quantum theory. For instance, we do not postulate a pilot-wave to guide the dynamics of the system, nor do we assume the system has definite position and velocity.}.

The Schr\"{o}dinger equation of the bipartite system is derived in Appendix A as
\begin{equation}
\label{jointSE}
    i\hbar\frac{\partial\Psi}{\partial t} = [-\frac{\hbar^2}{2m_a}\nabla_a^2 -\frac{\hbar^2}{2m_b}\nabla_b^2 + V_a + V_b]\Psi,
\end{equation}
where $\Psi({x}_a, {x}_b, t) = \sqrt{\rho({x}_a, {x}_b, t)}e^{iS(x_a, x_b, t)/\hbar}$. The separability of $\Psi$, depends on the separability of $\rho({x}_a, {x}_b, t)$, which in turn depends on the initial conditions (\ref{initCond})-(\ref{initCond2}). In Appendix B, we prove the following theorem:
\begin{Theorem} 
\label{theorem1}
Given the following two conditions
\begin{enumerate}
\item[1.] There is no interactions between the two subsystems $A$ and $B$, $V(x_a, x_b) = V_a(x_a) + V_b(x_b)$;
\item[2.] Locality of vacuum fluctuations. That is, the random fluctuations that $A$ and $B$ experience are independent from each other; 
\end{enumerate}
$\Psi({x}_a, {x}_b, t)$ is inseparable if and only if the initial joint probability density is inseparable, that is, (\ref{initCond}) holds.
\end{Theorem}
The locality of vacuum fluctuations is formally defined in (\ref{jointTP}). It has been utilized in Appendix A to prove (\ref{If3}). We explicitly include it in the theorem to emphasize that the inseparability of $\Psi({x}_a, {x}_b, t)$ can be maintained even if there is no interaction between $A$ and $B$ and their random fluctuations are independent\footnote{Theorem 1 assumes the bipartite system is in a pure quantum state described by $\Psi({x}_a, {x}_b, t)$. In a more general case, the bipartite system can be in a mixed state and described by a density matrix $\sigma_{ab}(t)$. In this case, separability is defined as that the bipartite density matrix is the probabilistic linear mixture of unentangled pure states~\cite{Hayashi15}. Given the linearity of Schr\"{o}dinger equation, Theorem 1 can be extended as that on the same two conditions, $\sigma_{ab}(t)$ is separable if and only if the initial density matrix $\sigma_{ab}(t_0)$ is separable.}.

The fact that $\Psi({x}_a, {x}_b, t)$ is inseparable as a pure state of a bipartite system implies it is an entanglement state~\cite{Nielsen}. We can define the entanglement measure as
\begin{equation}
    \mathcal{E} = 1 - Tr(\hat{\sigma}_a^2)
\end{equation}
Where $Tr(\sigma_a^2)$ is the purity of the reduced density matrix of subsystem $A$, $\sigma_a^2$. It can be computed using $\Psi(x_a, x_b, t)$ directly as
\begin{equation}
\label{purity}
\begin{split}
    Tr(\sigma_a^2) =& \int \Psi(x_a, x_b, t)\Psi^*(x'_a, x_b, t)\Psi(x_a, x'_b, t)\\
    &\times\Psi^*(x'_a, x'_b, t)dx_a dx_b dx'_a dx'_b.
\end{split}
\end{equation}
If $\Psi(x_a, x_b, t)=\psi(x_a, t)\phi(x_b, t)$, that is, a separable state, one can verify that $Tr(\sigma_a^2)=1$ and therefore $\mathcal{E}=0$. Otherwise, $Tr(\sigma_a^2)<1$ and $\mathcal{E}>0$.

\subsection{Three Aspects of Entanglement}
\label{EngtProc}
To clearly compare the difference of various conditions that result in either separable or entangled dynamics of the bipartite system, we summarize the discussions in the previous two subsections in Table 1. Several important observations can be made from this table.

\begin{table*}[ht]
\caption{Conditions for determining whether the dynamics of two subsystems are separable}
\label{tab:1}       % Give a unique label
\renewcommand*{\arraystretch}{1.4}
\begin{tabular}{|m{5cm}|m{1.5cm}|m{7cm}|m{1.7cm}|}
%\begin{tabu} to 0.8\textwidth { X[c] | X[c] | X[c] | X[c] }
\hline
 initial condition at $t_0$ & vacuum fluctuation & dynamics of the two subsystems separable? & classical or quantum? \\
\hline
$\rho(x_a,x_b, t_0)=\rho_a(x_a,t_0)\rho_b(x_b,t_0)$ & $I_f=0$ & $S_c=(S_c)_a+(S_c)_b$, separable dynamics& classical\\
\hline
$\rho(x_a,x_b, t_0)\ne \rho_a(x_a,t_0)\rho_b(x_b,t_0)$ & $I_f= 0$ & $S_c=(S_c)_a+(S_c)_b$, separable dynamics& classical\\
\hline
$\rho(x_a,x_b, t_0)=\rho_a(x_a,t_0)\rho_b(x_b,t_0)$ & $I_f\ne 0$ & $\Psi(x_a,x_b,t)=\psi_a(x_a,t)\psi_b(x_b,t)$, separable state& quantum\\
\hline
$\rho(x_a,x_b, t_0)\ne\rho_a(x_a,t_0)\rho_b(x_b,t_0)$ & $I_f\ne 0$ & $\Psi(x_a,x_b,t)\ne\psi_a(x_a,t)\psi_b(x_b,t)$, entangled state& quantum\\
\hline
\end{tabular}
\end{table*}

\begin{figure*}
\begin{center}
\includegraphics[scale=0.62]{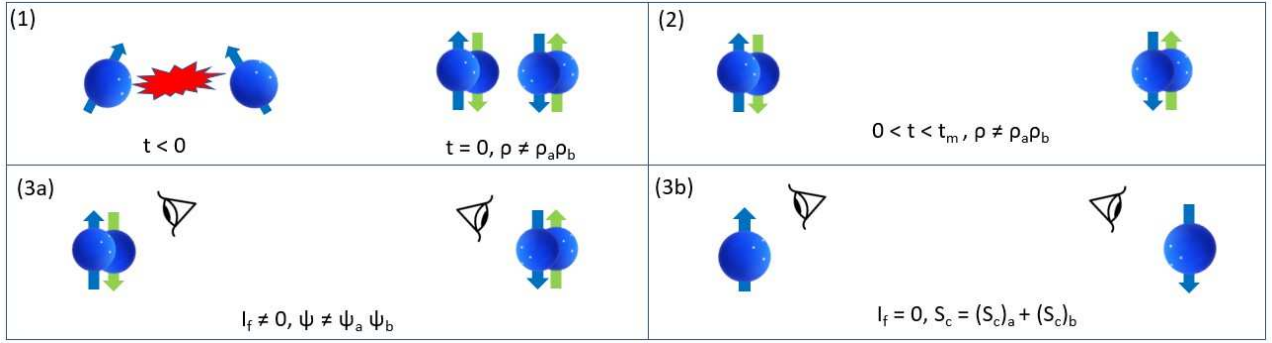}
\caption{Sketch of an entangled bipartite system. (1) Formation of inseparable correlation. There is physical interaction between two subsystems for a period of time at $t < 0$. Suppose at $t=0$, the physical interaction is turned off. Correlation between the two subsystems is established such that $\rho\ne\rho_a\rho_b$. Here the up and down arrows symbolize the correlated property, not necessarily related to spin. The composite system can be in a superposition state, represented by two close shapes. (2) Preservation of correlation. The two subsystems move apart from each other without physical interaction for $0<t<t_m$. The correlation encoded in $\rho\ne\rho_a\rho_b$ is preserved since the law of dynamics. (\ref{jointSE}) in the quantum mechanics case, admits solutions with $\rho\ne\rho_a\rho_b$. (3) Manifestation of correlation. In the quantum case (3a), before the actual measurement the wavefunction is inseparable. Measured results at time $t_m$ on the two subsystems are correlated. But in the classical case (3b) measurements will find no correlation since $I_f=0$ and $S_{C}=(S_C)_a+(S_C)_b$.}
\label{fig:1}       % Give a unique label
\end{center}
\end{figure*}

Table 1 shows that both conditions, $\rho(x_a,x_b, t_0)\ne\rho_a(x_a,t_0)\rho_b(x_b,t_0)$ and $I_f\ne 0$, must be true at the same time for the bipartite system to be in an entangled state. If there is no initial inseparable correlation exists between $A$ and $B$, subsequent time evolution will not introduce inseparable correlation since there is no interaction between $A$ and $B$ after $t_0$. Thus, a pre-requisite for $A$ and $B$ to be entangled is that there must an initial inseparable correlation between them at $t_0$, and this is encoded in the condition $\rho(x_a,x_b, t_0)\ne\rho_a(x_a,t_0)\rho_b(x_b,t_0)$. But this condition alone is not sufficient. If the term $I_f$ vanishes, the least observability principle becomes the least action principle, and the inseparable correlation encoded in $\rho({x}_a, {x}_b, t)$ bears no consequence since $S_c$ itself is still separable due to the law of marginal probability (\ref{lmp}). The dynamics of $A$ and $B$ are still independent, which is the case in classical mechanics. But because of the existence of $I_f$, the impact of the inseparability of $\rho({x}_a, {x}_b, t)$ is manifested, resulting in the entanglement effect. Therefore, one can argue that $I_f$ is the source for manifesting entanglement. But $I_f$, which is the cumulative relative entropy produced by vacuum fluctuations, represents purely an information requirement. Its role of manifesting entanglement is achieved even though the two subsystems are moving apart and without interaction, and the vacuum fluctuations of the two subsystems are independent. One may insist that there must be a hidden physical mechanism to keep $\rho({x}_a, {x}_b, t)$ inseparable when two subsystems move far apart without explicit interaction between them, and such a hidden physical mechanism must be non-local. Such an insistence is unfounded. The inseparable correlation between the two subsystems encoded in $\rho({x}_a, {x}_b, t)$ is purely informational. Entanglement is just a new way of manifesting the correlation that is not seen in classical mechanics due to the existence of $I_f$. 

These observations and analysis shows that the process for quantum entanglement can be deciphered into three aspects. First, the formation of inseparable correlation. The correlation between the two subsystems are initially caused by physical interaction between them at $t<0$ and encoded by the inseparability of $\rho({x}_a, {x}_b, t=0)$. Second, the preservation of such correlation. Schr\"{o}dinger equation (\ref{jointSE}) allows such correlation to be preserved in subsequent dynamics even though there is no further interaction. Actually the inseparability of $\rho({x}_a, {x}_b, t)$ is true in both classical and quantum mechanics. However, the inseparability of $\rho({x}_a, {x}_b, t)$ bears no consequence in classical mechanics, since $S_c$ is still separable, as shown in (\ref{SC1}). Third, 
{the manifestation of entanglement effect}. In quantum mechanics, the effect of inseparability of $\rho({x}_a, {x}_b, t)$ is manifested through $I_f$ because without $I_f$ the dynamics of the composite system just follows the laws of classical mechanics. Because of the existence of $I_f$, the wave function of the bipartite system is inseparable, even though with no interaction, independent vacuum fluctuations. Subsequent measurement either $A$ or $B$ just confirms such an inseparable correlation between $A$ and $B$~\cite{Hensen}. The first and second aspects are the same for both classical and quantum mechanics, while the third aspect is unique to quantum mechanics. Fig.1 depicts these three aspects of quantum entanglement.

To summarize, entanglement is an inseparable correlation encoded in previous physical interaction between the two subsystems. It is causal in this sense at $t<0$. Such correlation is preserved in later dynamics even if there is no further interaction, and with independent vacuum fluctuations, of the two subsystems. The effect of inseparable correlation is then manifested in the case of quantum mechanics through the requirement to extremize an informational metric $I_f$. The preservation and manifestation aspects of entanglement are non-causal. 

The discussions in this section are rather abstract. We will give two concrete examples in the next section to illustrate the ideas discussed here.

\section{Examples of Two Entangled Free Particles} 
\label{sec:examples}
\subsection{Superposition of Energy Eigen States}
Consider two free particles $A$ and $B$ with initial joint probability distribution $\rho(x_a, x_b, 0)$. Suppose there is no more interaction between them after $t_0=0$ so that the time evolution of wave function in (B1) becomes
\begin{equation}
\label{jointSE3}
    \Psi(x_a, x_b, t)=e^{-i(\hat{H}_a+\hat{H}_b)t/\hbar}\Psi(x_a, x_b, 0).
\end{equation}
However, they may have been interacting before $t_0$ such that the initial condition $\rho(x_a, x_b, 0)$ is inseparable. Suppose we choose
\begin{align}
\label{initP}
    &\rho(x_a, x_b, 0)=\frac{1}{Z}(1+\cos\theta), \text{   and}\\
   &\theta =  \frac{1}{\hbar}(p_{a1}x_a+p_{b1}x_b - p_{a2}x_a+p_{b2}x_b),
\end{align}
where $Z$ is a normalization factor, $\{p_{a1}, p_{a2}\}$ are two different momentum values for particle $A$, and $\{p_{b1}, p_{b2}\}$ for particle $B$. This is equivalent to choose the initial wave function
\begin{equation}
    \Psi(x_a, x_b, 0) = \frac{1}{\sqrt{2Z}}\{e^{\frac{i}{\hbar}(p_{a1}x_a+p_{b1}x_b)}+e^{\frac{i}{\hbar}(p_{a2}x_a+p_{b2}x_b)}\}.
\end{equation}
Inserting $\Psi(x_a, x_b, 0)$ into (\ref{jointSE3}), we get
\begin{equation}
\begin{split}
    \Psi(x_a, x_b, t) =& \frac{1}{\sqrt{2Z}}\{e^{\frac{i}{\hbar}(p_{a1}x_a+p_{b1}x_b-E_{a1}t-E_{b1}t)} \\
    &+e^{\frac{i}{\hbar}(p_{a2}x_a+p_{b2}x_b-E_{a2}t-E_{b2}t)}\},
\end{split}
\end{equation}
where $E_{jk}=p^2_{jk}/m_j$ and $j\in\{a,b\}$, $k\in\{1,2\}$. $\Psi(x_a, x_b, t)$ is an entangled state. When measuring $A$ in energy eigen state with $E_{a1}$, $B$ is in its energy eigen state with $E_{b1}$, and when measuring $A$ in energy eigen state with $E_{a2}$, $B$ is in its energy eigen state with $E_{b2}$. Suppose we further choose $E_{a1}+E_{b1}=E_{a2}+E_{b2} = E$, $\Psi(x_a, x_b, t)$ is then simplified to
\begin{equation}
\label{jointSE4}
\begin{split}
    \Psi(x_a, x_b, t) =& e^{-\frac{i}{\hbar}Et}\Psi(x_a, x_b, 0).
\end{split}
\end{equation}
This implies $\rho(x_a, x_b, t)=\rho(x_a, x_b, 0)$ and independent of time. The entanglement between the two particles over time stays the same as the initial entanglement. This can be verified by inserting (\ref{jointSE4}) into (\ref{purity}) to compute the purity of density matrix $\sigma_a$
\begin{equation}
\label{purity2}
\begin{split}
    &Tr(\sigma_a^2(t)) = \int \Psi(x_a, x_b, 0)\Psi^*(x'_a, x_b, 0)\\
    &\times\Psi(x_a, x'_b, 0)\Psi^*(x'_a, x'_b, 0)dx_a dx_b dx'_a dx'_b.
\end{split}
\end{equation}
Since $Tr(\sigma_a^2(t))$ is independent of $t$, so is the entanglement measure $\mathcal{E}=1-Tr(\sigma_a^2(t))$. The inseparability of the initial condition $\Psi(x_a, x_b, 0)$ ensures $\mathcal{E}>0$. Thus, it is a constant over time.
If we further choose $\{p_{a1}, p_{a2}\}$ positive and $\{p_{a1}, p_{a2}\}$ negative, the two particles are moved along opposite directions, but the entanglement measure $\mathcal{E}$ stays the same. 

However, the probability density distribution (\ref{initP}) is a cosine function over the entire one dimensional space. The probability of finding a particle in a particular position is unchanged even though the two particles move away from each other. Furthermore, the probability density distribution (\ref{initP}) cannot be normalized properly. This simple example cannot be physically realized. It only serves as a heuristic example for Theorem 1. It shows that the entanglement is due to the initial condition (\ref{initP}), and manifested by $I_f$ even though there is no interaction between the two particles.  

\begin{figure*}
\begin{center}
\includegraphics[scale=1.5]{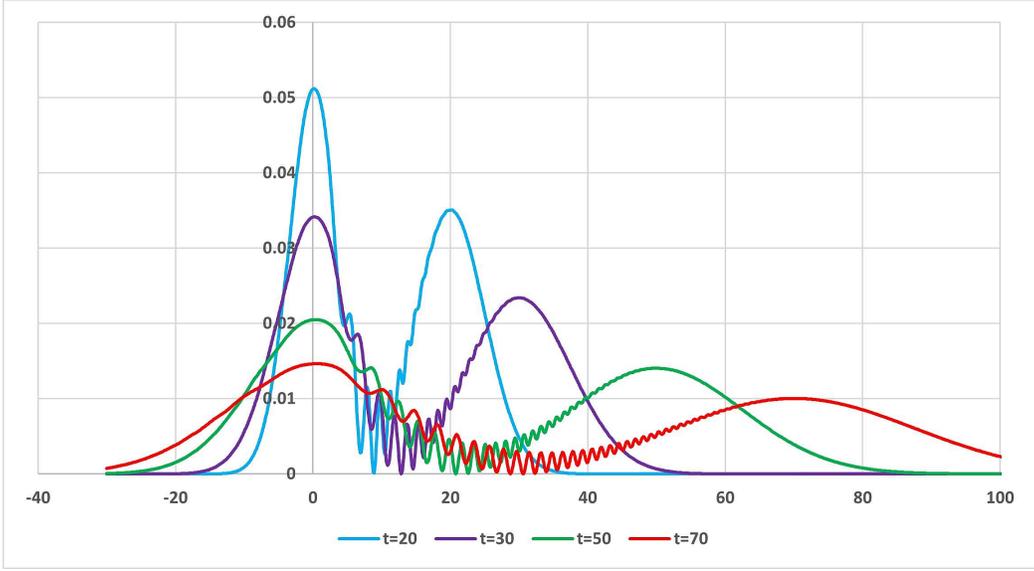}
\caption{Time evolution of the joint probability density distribution, defined in (\ref{EGRho}), for two entangled Gaussian wave packets. The plots are obtained by setting the parameters $m_b=2m_a$, $\alpha_b=\alpha_a/\sqrt{2}$, such that $\tau_a=\tau_b=\tau$, and $\alpha_ap_0 = 2\sqrt{2}, \alpha_a\hbar=1/2\sqrt{2}$. Time parameter $t$ is set in units of $\tau$. For instance, $t=20$ means $t=20\tau$. The plots show that Gaussian wave packet $A$ is moving to the right and away from wave packet $B$ which peak stays in the original position. Both wave packets show dispersion effects as time evolves. The oscillations in the overlapped regions reflect the interference effect, which indicates entanglements.}
\label{fig:2}       % Give a unique label
\end{center}
\end{figure*}

\subsection{Entangled Gaussian Wave Packets}
\label{sec:Gaussian}
A more realistic and physically realizable example is when a free particle is described by a Gaussian wave packet. A free particle $A$ can be described by a Gaussian wave packet with initial momentum distribution~\cite{Robinett}
\begin{equation}
    \label{GMF}
    \phi^A_G(p_a, t_0)=\sqrt{\frac{\alpha_a}{\sqrt{\pi}}}e^{-\alpha_a^2(p_a-p_0)^2/2},
\end{equation}
where $p_0$ is the initial central momentum value and assuming the initial central position $x_0=0$. The variance of momentum is given by $\Delta p^a_0=1/\sqrt{2}\alpha_a$. The time-dependent wave function for such Gaussian wave packet is derived from (\ref{jointSE3}) and given by~\cite{Robinett}
\begin{equation}
    \label{GWF}
    \begin{split}    
    \psi^A_G(x_a, t)&=\frac{1}{\sqrt{\sqrt{\pi}\alpha_a\hbar(1+it/\tau_a)}}e^{i(p_0x_a-E_0t)/\hbar}\\
    &\times e^{-(x_a-p^a_0t/m)^2/2(\alpha_a\hbar)^2(1+it/\tau_a)},
    \end{split}
\end{equation}
where $E_0=p_0^2/2m_a$, and $\tau_a=m_a\hbar\alpha_a^2$ is the spreading time. This gives the probability density 
\begin{equation}
    \label{Grho}
    \rho^A_G(x_a, t) = \frac{1}{\sqrt{\pi}\beta^a_t}e^{-(x_a-p_0t/m)^2/(\beta^a_t)^2},
\end{equation}
where $\beta^a_t=\beta_a\sqrt{1+(t/\tau_a)^2}$ and $\beta_a=\alpha_a\hbar$. $\psi^A_G(x,t)$ describes a free particle $A$ with initial central position $x_0=0$ and moving with initial momentum value $p^a_0$. Suppose there is another free particle with initial central position $x_0=0$ and initial momentum value $p_0=0$. The wave packet is then given by
\begin{equation}
    \label{GWF}  
    \psi^B_G(x_b, t)=\frac{1}{\sqrt{\sqrt{\pi}\alpha_b\hbar(1+it/\tau_b)}}e^{-x_b^2/2(\alpha_b\hbar)^2(1+it/\tau_b)},
\end{equation}
with $\tau_b=m_b\hbar\alpha_b^2$. Denote $\beta^b_t=\beta_b\sqrt{1+(t/\tau_b)^2}$, the corresponding probability density is given by
\begin{equation}
    \label{Grho2}
    \rho^B_G(x_b, t) = \frac{1}{\sqrt{\pi}\beta^b_t}e^{-x_b^2/(\beta_t^b)^2}.
\end{equation}
We now wish to construct an entangled bipartite system with the two free particles $A$ and $B$. Suppose we choose the following initial joint probability distribution
\begin{equation}
    \label{EGRho0}
    \begin{split}  
    \rho(x_a,x_b,0)=&\frac{N^2}{2}[\rho_G^A(x_a,0)+\rho_G^B(x_b,0)+2\sqrt{\rho_G^A\rho_G^B}\cos\theta_0]\\
    =&\frac{N^2}{2}[\frac{1}{\sqrt{\pi}\beta_a}e^{-x_a^2/\beta_a^2}+\frac{1}{\sqrt{\pi}\beta_b}e^{-x_b^2/\beta_b^2} \\
    &+\frac{\cos(p_0x_a/\hbar)}{\sqrt{\pi\beta_a\beta_b}}e^{-x_a^2/2\beta_a^2-x_b^2/2\beta_b^2}].
    \end{split}
\end{equation}
This is equivalent to choose the wave function for the bipartite system $A$ and $B$ as~\cite{Robinett}
\begin{equation}
    \label{EGWP}
    \Psi(x_a, x_b, t)=\frac{N}{\sqrt{2}}[\psi^A_G(x_a,t)+\psi^B_G(x_b,t)],
\end{equation}
where $N=1/\sqrt{1+e^{-(\alpha_a p_0/2)^2}}$ is a normalization factor \cite{Robinett}. One can verify $\Psi(x_a, x_b, t)$ is a solution of Schr\"{o}dinger equation of two free particles. From (\ref{EGWP}), the joint probability density can be calculated as
\begin{equation}
\label{EGRho}
    \rho(x_a,x_b,t)=\frac{N^2}{2}[\rho_G^A(x_a,t)+\rho_G^B(x_b,t)+2\sqrt{\rho_G^A\rho_G^B}\cos\theta_t],
\end{equation}
where
\begin{equation}
    \label{thetaT}
    \theta_t=\frac{1}{\hbar}(p_0x_a-E_0t)+\frac{(x_a-p_0t/m_a)^2t}{2\tau_a(\beta_t^a)^2}-\frac{x_b^2t}{2\tau_b(\beta_t^b)^2}.
\end{equation}
Since Eq.(\ref{EGRho0}) cannot be written in the form of (\ref{initCond2}) and $\Psi(x_a, x_b, t)$ is a pure state, by Theorem 1, such bipartite Gaussian wave packets are entangled. They stay entangled even though they are moving away from each other and there is no interaction between the two subsystems. Fig. 2 depicts how $\rho(x_a, x_b, t)$ evolves as particle $B$ moves away from $A$.

\begin{figure*}
\begin{center}
\includegraphics[scale=1.5]{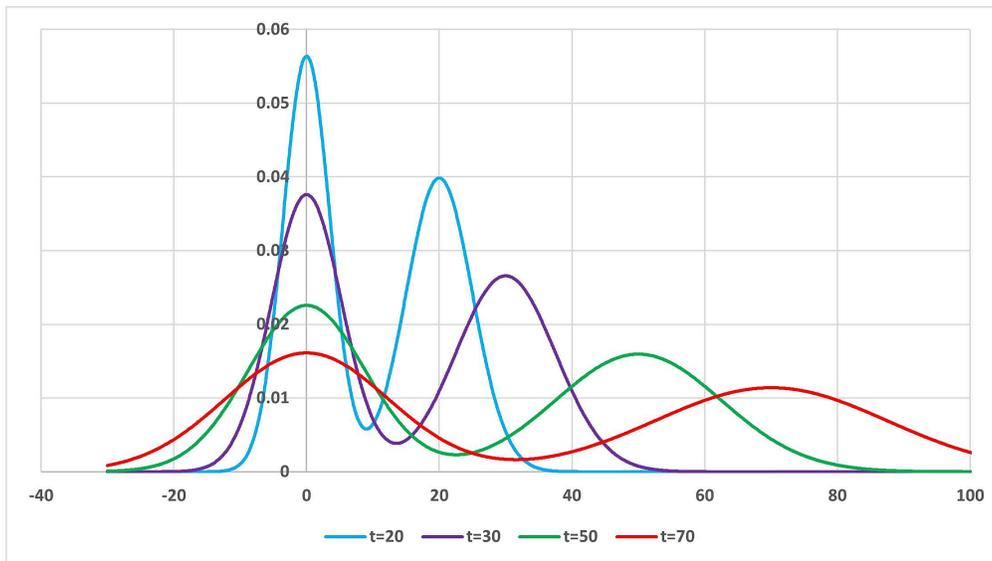}
\caption{Time evolution of the joint probability density distribution, defined in (\ref{EGRho}) but without the third term, for two un-entangled Gaussian wave packets. This is corresponding to the mixed state (\ref{mixedState}). The parameters are set exactly the same as those in Fig. 2. Both wave packets still show dispersion effects as time evolves, but there is no interference effect in the overlapped region, indicating no entanglement.}
\label{fig:2}       % Give a unique label
\end{center}
\end{figure*}

Let's take a closer look at (\ref{EGRho}). Without the third term, $\rho(x_a,x_b,t)$ is essentially corresponding to a classical mixture of two Gaussian distributions with density matrix
\begin{equation}
    \label{mixedState}
    \sigma_{ab}(t)=\frac{1}{2}(|\psi_G^A\rangle\langle\psi_G^A|\otimes I_B + I_A\otimes|\psi_G^B\rangle\langle\psi_G^B|).
\end{equation}
There is no entanglement between them. The third term in (\ref{EGRho}) is reflecting the correlation between two wave packets and therefore responsible for the entanglement. It depends on the overlap of the two Gaussian wave packets. This is visually seen in Fig. 2 as the oscillations in the overlapped region of the probability distributions. One may think that as the two Gaussian wave packets move apart further and further, the overlap will become smaller and smaller, so that the entanglement will disappear eventually. However, there are two effects that are unique in quantum mechanics and contribute to preserving the correlation even though the two wave packets move apart. First, the dispersion of the wave packet. As the two Gaussian wave packets move away from each other, the shapes of their probability distribution also spread as time evolves. If the speed of the wave packet spreading catches the speed of the wave packet movement, the overlap between the two wave packets does not reduce over time. Second, the interference in the overlapped region also propagates as time evolves. This can be seen in Fig. 2 that the oscillations in the overlapped region spread together with the wave packets. Both the dispersion and interference propagation effects maintain the inseparable correlation, hence the entanglement, between the two wave packets even though they are moving apart. The actual entanglement measure can be calculated by inserting (\ref{EGWP}) into (\ref{purity}). 

The dispersion of the Gaussian wave packet is a quantum phenomenon. In the classical limit, the Gaussian wave packet moves with a determined speed, so that $\Delta p_0 \to 0$ and $\alpha\to\infty$. Meanwhile, $\hbar\to 0$ but we require $\beta=\alpha\hbar$ as a constant so that $\beta_t=\beta$. In such a limit, the probability density defined (\ref{Grho}) will not change the shape while it moves. However, dispersion effect alone does not warrant entanglement. This can be seen in Fig. 3, which depicts the probability distribution $\rho(x_a, x_b, t)$ in (\ref{EGRho}) without the third term, and corresponds to a mixed state represented in (\ref{mixedState}). The dispersion effect is still there, but there is no entanglement. Thus, the interference effect is necessary to manifest entanglement. The interference effect is due to the inseparable initial condition of the joint probability density (\ref{EGRho0}), in accordance with Theorem 1.

But what causes the dispersion of the Gaussian wave packet and the propagation of the interference effect shown in Fig. 2? As discussed in Section \ref{EngtProc}, it must be caused by the information metrics $I_f$. The existence of $I_f$ means that while the Gaussian wave packet is moving, it must also change its shape such that the observable information $I_f$ is minimized together with the classical observation information. This posts an additional constraint on how the two Gaussian wave packets should behave as time evolves. As a consequence, the initial inseparable correlation between the two Gaussian wave packets are preserved and manifested as an entanglement effect even though the two wave packets are moving apart. The fundamental question here is whether the requirement to minimize $I_f$ implies the existence of any non-local causal connection. This is the subject in the next section.

\section{Discussion and conclusions} 
\label{sec:discussion}

\subsection{Non-locality and Causality}
In the example of two entangled particles represented by Gaussian wave packets, we argue that the initial entanglement between the two particles is preserved and manifested by the information metrics $I_f$ even when the two particles are moving away from each other without interaction. Recall that $I_f$ is defined to measure the information distance between probability distribution due to vacuum fluctuations and probability distribution without vacuum fluctuations. Importantly, such vacuum fluctuations are assumed to be local when we derive the Schr\"{o}dinger equation in Appendix A. In other words, even though there is no interaction between the two particles, and the vacuum fluctuation of one particle has no impact on the other one, the initial entanglement is still preserved and propagated due to such local vacuum fluctuation, via the requirement to minimize information metrics $I_f$ together with observable information from classical dynamics. The process does not involve non-local interaction after all. After sufficient amounts of time evolution, the two particles may be spatially far apart, but the inseparable correlation is preserved. And because the inseparable correlation is preserved, when properties on particle $A$ are measured, one can predict properties of particle $B$. Subsequent measurement on particle $A$ does not cause any impact on $B$. Instead, the measurement just reveals the inseparable correlation preserved due to $I_f$ via local vacuum fluctuations. The correlations between measurement outcomes between two particles appear non-local, but there is no non-local causal effect between them. 

Imposing the constraint of extremizing information metrics $I_f$ is purely an information requirement. $I_f$ itself is inseparable if the initial joint probability is inseparable. Thus, $I_f$ is an inseparable global metric even though the underlying mechanics of vacuum fluctuation is local. This is similar to the second law of thermodynamics where the entropy of, for instance, an ensemble of isolated ideal molecule gas, is maximized even though the underlying mechanics is just local collisions among molecules. The key point here is that extremizing a global information metric does not require non-local interaction as an underlying mechanism. Viewing the underlying mechanism of entanglement phenomenon from an information perspective can dismiss the mystery of ``spooky action at a distance". 

It is also worth pointing out that variation of $I_f$ over the probability density gives a term similar to the Bohm potential, as shown in (\ref{QHJ}). Therefore, mathematically, the dispersion and interference effects discussed in the entangled Gaussian wave packets can also be explained by the Bohm potential. However, the non-local nature of the Bohm potential still leaves speculation of some sorts of non-local interaction as the underlying mechanism. On the other hand, with the principle of least observability presented here, $I_f$ is just an information metric on local vacuum fluctuations. There is no room to speculate non-local interaction as the underlying mechanism. We believe this advantage is conceptually fundamental.

\subsection{Revisiting the EPR Thought Experiment}
Given the new insights on entanglement discussed in the previous subsection, it is worth revisiting the EPR thought experiment and confirm that there is no paradox.  

Suppose an entangled pair of particle $A$ and $B$ are remotely separated and there is no interaction between them, just like the two entangled Gaussian wave packets discussed earlier. Alice is moving with the same speed $p_0$ with $A$ while Bob is staying at the same original with $B$. Initially both Alice and Bob share the same information of the wave function $\Psi(x_a, x_b, t)$ as defined in (\ref{EGWP}). Now suppose Alice performs a measurement on $A$ and obtain momentum eigen value $p_a$. From Alice's point of view, the wave function is updated to some momentum eigen state $\Psi'(x_a, x_b, t)=\psi^A_p(x_a,t)\psi^B_p(x_a,t)$ and she can predict that if Bob perform a measurement on $B$, he will obtain some momentum eigen value $p_b$. Suppose Alice does not send the measurement outcome to Bob. Given there is no interaction between $A$ and $B$, the vacuum fluctuation is local, and the measurement action of Alice is also local. There is no direct influence on particle $B$. Thus, from Bob's point of view, he still describes both $A$ and $B$ using the original wavefunction $\Psi(x_a, x_b, t)$. Alice's and Bob's descriptions on $B$ are different but both are valid. This is the relational quantum mechanics (RQM) interpretation advocated by Rovelli~\cite{Rovelli:1995fv}. The RQM interpretation is compatible with the least observability principle since as we show in earlier sections, the preserving and manifesting of quantum entanglement is due to the existence of information metrics $I_f$. There is no non-local interaction underlying the origin of $I_f$ for the bipartite system $A$ and $B$. Thus, it is legitimate for Bob to retain the original wave function as his description of $A$ and $B$, even Alice has performed a measurement on $A$ remotely. Of course, if Alice sends the measurement results to Bob, Bob's observable information on $A$ and $B$ is updated, consequently Bob's description on the bipartite system is changed.

%Once one admits that the measurement outcome is observer dependent, and the assignment of quantum state to a system is also observer dependent, several famous seemingly paradoxes can be resolved. These include the EPR thought experiment~\cite{Rovelli07, Yang2018}, Wigner's friend thought experiment, and its extended version recently discussed intensively~\cite{FR2018, Yang2019}. 
In the original EPR thought experiment, when one is able to predict with certainty the value of a physical quantity without disturbing a system, it is said that there exists an element of physical reality corresponding to the physical quantity. However, since the measurement outcome on the momentum value of $A$ is observer dependent, the prediction of momentum value of $B$ is also observer dependent, or context dependent. There is no absolute element of physical reality without specifying the measurement context. If Bob does not know Alice's measurement outcome and does not perform his own measurement based on Alice's measurement outcome, the physical reality Alice perceived on $B$ is only valid to Alice, since from Bob's perspective, nothing happens to $B$. The key here is that the inseparable correlation between the two particles is a non-causal correlation. The least observability principle presented in this paper gives the actual mathematical formulation to confirm that the preservation and manifestation of such inseparable correlation is a non-causal, through the introduction of information metrics $I_f$. Such mathematical confirmation is complementary to the conceptual effort of resolving the EPR paradox using RQM interpretation~\cite{Rovelli07, Yang2018}. Now suppose instead Alice performs measurement on the position value of $A$ and is able to predict the position of $B$. However, again such prediction does not imply there exists absolute physical reality of position for $B$, unless Bob knows Alice's result and performs corresponding measurement. Alice's ability to predict both momentum and position of $B$ does not imply there exists absolute physical reality of momentum and position of $B$ in the same time. Thus, the claim from the original EPR paper that quantum mechanics is incomplete is dismissed. 

\subsection{On the Bohm Quantum Potential}
Eq. (\ref{BohmQ}) is mathematically identical with the Bohm quantum potential for a bipartite quantum system. In Bohmian quantum mechanics, this potential is postulated, and is associated with mysterious non-local hidden mechanism. But in our theory, it is simply a result of variation of the information metrics $I_f$ in (\ref{If3}) with respect to the probability density. There is no need to postulate a pilot-wave to guide the dynamics of the system. From the perspective of least observability principle, the ``guiding wave" is essentially equivalent to the requirement to minimize the information metrics $I_f$ during the dynamics of the system. Eq. (\ref{BohmQ}) generally cannot be separated into two individual terms for the two subsystem respectively. But the reason of this inseparability is due to the initial condition (\ref{initCond}). Furthermore, such an inseparability is preserved and propagated through the vacuum fluctuations of the bipartite system. Locality of the vacuum fluctuations rules out the need for a non-local mechanism. In summary, a mathematical term identical to the Bohm quantum potential is derived from the information metrics $I_f$, and our theory shows that the inseparability of this term is not associated with a non-local mechanism.

\subsection{Limitations}
Actual experiments to confirm Bell theorem~\cite{Hensen} are typically not based on entanglement of momentum or position, but based on other degrees of freedom such as electron spins, or polarization of photons. In order to explain the entanglement of such degrees of freedom from an information perspective, we need to define information metrics in addition to $I_f$ for these degrees of freedom, then apply the least observability principle to derive quantum dynamics equations, for example, the Dirac equation for electron spins. If these metrics just measure information due to local dynamics, and the inseparable correlation of spins of a bipartite system are shown to be preserved and manifested because of these information metrics, it then confirms that there is no-local causal effect in the entanglement phenomenon of spins. This is beyond the current scope of this paper. However, it is an interesting future research topic. 

\subsection{Conclusions}
Formulating quantum mechanics based on the extended least action principle, or the least observability principle, helps us to gain new insights on quantum entanglement. This is because the principle clearly demonstrates how classical mechanics becomes quantum mechanics from the information perspective. In the paper, we investigate the mechanism of entanglement of two free quantum systems. Mathematically, we show that for such a bipartite system, two conditions are both necessary in order to exhibit entanglement behavior. First, the initial joint probability distribution must be inseparable. Second, there is a requirement to extremize the information metrics $I_f$ due to local vacuum fluctuations. Both conditions must be met at the same time. Without the first condition, a bipartite quantum system is described by a separable state and exhibits no entanglement correlation. On the other hand, without the second condition, the bipartite system follows classical mechanics. Even if the initial joint probability distribution is inseparable, there is no impact on the dynamics of each subsystem, thanks to the validity of the law of marginal probability. 

Thus, entanglement of two spatially separated quantum subsystems can be understood as the following. The initial inseparable correlation between them is preserved and manifested due to the constraint to extremize the information metrics $I_f$. $I_f$ captures the information distance between probability distribution with vacuum fluctuations to probability distribution without vacuum fluctuations, and such vacuum fluctuations are local. The conclusion is that entanglement is an inseparable correlation but at the same time also a non-causal correlation. Manifesting of the entanglement relation does not require non-local causal effect. This is illustrated in Section \ref{sec:Gaussian} with two entangled Gaussian wave packets.

The fact that the information metrics $I_f$ is responsible for the quantumness has other implications. For instance, our theory shows that the Bohm quantum potential originates from extremizing $I_f$. Historically, the Bohm potential is widely considered as non-local. The non-locality of the Bohm potential is a mystery and one would naturally speculate there is a hidden non-local mechanism underlying it. However, we show that the Bohm potential originates from an informational requirement which does not need a non-local causal mechanism. This is similar to the second law of thermodynamics statistics where a global information quantity, entropy of, for instance, an ensemble of ideal molecule gas, is maximized but the under mechanism is just local collisions among molecules.

\onecolumngrid

\pagebreak
%========================================

\appendix

\section{Derivation of Schr\"{o}dinger Equation for a Bipartite System}
We assume that from $t \ge 0$, there is no interaction between the two subsystems of the bipartite system. It is also natural to assume that the vacuum fluctuations for each subsystem are independent from each other. This implies that the vacuum fluctuations are local, and the transition probability of the composite system is separable, as shown in (\ref{jointTP}). Next we extend the definition of $I_f$ in (\ref{DLDivergence}) to the bipartite system:
\begin{align*}
\label{DLDivergencefor2}
    I_f &=: \sum_{j=0}^{N-1}E_{{w}}[D_{KL}(\rho ({x}_a,{x}_b, t_j) || \rho ({x}_a+{w}_a, {x}_b+{w}_b, t_j)] \\
    &=\sum_{j=0}^{N-1}\int d{w}_ad{w}_b d{x}_ad{x}_b\wp({x}_a+{w}_a, {x}_b+{w}_b, t_j|{x}_a,{x}_b, t_j)\rho ({x}_a,{x}_b, t_j)ln \frac{\rho ({x}_a,{x}_b, t_j)}{\rho ({x}_a+{w}_a, {x}_b+{w}_b, t_j)} \\
    &=\sum_{j=0}^{N-1}\int d{w}_ad{w}_b d{x}_ad{x}_b\wp_a({x}_a+{w}_a, t_j|{x}_a, t_j)\wp_b({x}_b+{w}_b, t_j|{x}_b, t_j)\rho ({x}_a,{x}_b, t_j)ln \frac{\rho ({x}_a,{x}_b, t_j)}{\rho ({x}_a+{w}_a, {x}_b+{w}_b, t_j)}.
\end{align*}
Expanding the logarithm function %similarly to (\ref{Taylor2}),
\begin{align}
    ln\frac{\rho ({x}_a,{x}_b, t_j)}{\rho ({x}_a+{w}_a, {x}_b+{w}_b, t_j) } = & \frac{1}{\rho}\sum_i\partial_{ia}\rho w_{ia} + \frac{1}{2\rho} \sum_{i}\partial_{ia}^2\rho w_{ia}^2 -\frac{1}{2}(\frac{1}{\rho}\sum_i\partial_{ia}\rho w_{ia})^2 \\
    &+\frac{1}{\rho}\sum_i\partial_{ib}\rho w_{ib} + \frac{1}{2\rho} \sum_{i}\partial_{ib}^2\rho w_{ib}^2 -\frac{1}{2}(\frac{1}{\rho}\sum_i\partial_{ib}\rho w_{ib})^2.
\end{align}
We have
\begin{align*}
    I_f =& -\sum_{j=0}^{N-1}\int d{w}_ad{w}_b d{x}_ad{x}_b\wp_1\wp_2 [\sum_i\partial_{ia}\rho w_{ia} + \frac{1}{2} \sum_{i}\partial_{ia}^2\rho w_{ia}^2 -\frac{1}{2\rho}(\sum_i\partial_{ia}\rho w_{ia})^2 \\
    &+ \sum_i\partial_{ib}\rho w_{ib} + \frac{1}{2} \sum_{i}\partial_{ib}^2\rho w_{ib}^2 -\frac{1}{2\rho}(\sum_i\partial_{ib}\rho w_{ib})^2] \\
    =&\frac{1}{2}\sum_{j=0}^{N-1}\int d{x}_ad{x}_b\sum_i\{\langle w_{ia}^2\rangle [\frac{(\partial_{ia}\rho)^2}{\rho} - \partial_{ia}^2\rho] + \langle w_{ib}^2\rangle [\frac{(\partial_{ib}\rho)^2}{\rho} - \partial_{ib}^2\rho]\}
\end{align*}
The last step uses the fact that $\langle w_{ia}\rangle = \langle w_{ib}\rangle = 0$. Taking the assumption that $\rho$ is a regular function and its gradient with respect to ${x}_a$ or ${x}_b$ approaches zero when $|{x}_a|, |{x}_b|\to\pm\infty$, we have
\begin{equation*}
    I_f = \frac{1}{2}\sum_{j=0}^{N-1}\int d{x}_ad{x}_b\sum_i\{\langle w_{ia}^2\rangle \frac{(\partial_{ia}\rho)^2}{\rho} + \langle w_{ib}^2\rangle\frac{(\partial_{ib}\rho)^2}{\rho}\}.
\end{equation*}
Substituting $\langle w_{ia}^2\rangle = \hbar\Delta t/2m_a$ and $\langle w_{ib}^2\rangle = \hbar\Delta t/2m_b$, and taking $\Delta t\to 0$, we get
\begin{equation}
\label{If2}
    I_f = \int d{x}_ad{x}_bdt\{\frac{\hbar}{4m_a}\frac{\nabla_a\rho\cdot\nabla_a\rho}{\rho} + \frac{\hbar}{4m_b}\frac{\nabla_b\rho\cdot\nabla_b\rho}{\rho}\}.
\end{equation}
Combined with (\ref{cAction2}), the total amount of observable information for the bipartite system is
\begin{equation}
\label{totalI2}
    I = \frac{2}{\hbar}\int d{x}_ad{x}_bdt\rho\{\frac{\partial S}{\partial t} + \frac{1}{2m_a}\nabla_a S\cdot\nabla_a S
    + \frac{1}{2m_b}\nabla_b S\cdot\nabla_b S + V_a+V_b+\frac{\hbar^2}{8m_a}\frac{\nabla_a\rho\cdot\nabla_a\rho}{\rho^2} + \frac{\hbar^2}{8m_b}\frac{\nabla_b\rho\cdot\nabla_b\rho}{\rho^2}\}
\end{equation}
Variations with respect to $S$ and $\rho$, respectively, give two equations,
\begin{align}
    &\frac{\partial \rho}{\partial t} + \frac{1}{m_a}\nabla_a\cdot(\rho\nabla_a S)
    + \frac{1}{m_b}\nabla_b\cdot(\rho\nabla_b S) = 0; \\
    &\frac{\partial S}{\partial t} + \frac{1}{2m_a}\nabla_a S\cdot\nabla_a S
    + \frac{1}{2m_b}\nabla_b S\cdot\nabla_b S + V_a+V_b - \frac{\hbar^2}{2m_a}\frac{\nabla_a^2\sqrt{\rho}}{\sqrt{\rho}} - \frac{\hbar^2}{2m_b}\frac{\nabla_b^2\sqrt{\rho}}{\sqrt{\rho}} = 0.
\end{align}
Defined $\Psi({x}_a, {x}_b, t) = \sqrt{\rho({x}_a, {x}_b, t)}e^{iS/\hbar}$, it can be verified that that the two equations above are equivalent to the Schr\"{o}dinger equation in (\ref{jointSE}). 

Equation (\ref{If2}) shows that $I_f$ is inseparable since $\rho({x}_a, {x}_b, t) \ne  \rho_a({x}_a, t)\rho_b({x}_b, t)$. On the other hand, suppose $\rho({x}_a, {x}_b, t) =  \rho_a({x}_a, t)\rho_b({x}_b, t)$, then $\nabla_a \rho = \rho_b\nabla_a\rho_a$. Similarly, $\nabla_b \rho = \rho_a\nabla_b\rho_b$, then
\begin{align}
    I_f &= \int d{x}_ad{x}_bdt\{\frac{\hbar}{4m_a}\frac{\nabla_a\rho_a\cdot\nabla_a\rho_a}{\rho_a}\rho_b + \frac{\hbar}{4m_b}\frac{\nabla_b\rho_b\cdot\nabla_b\rho_b}{\rho_b}\rho_a\} \\
    &= \frac{\hbar}{4m_a}\int d{x}_adt\frac{\nabla_a\rho_a\cdot\nabla_a\rho_a}{\rho_a} + \frac{\hbar}{4m_b}\int d{x}_bdt\frac{\nabla_b\rho_b\cdot\nabla_b\rho_b}{\rho_b} = (I_f)_a + (I_f)_b,
\end{align}
and is clearly separable into two independent terms, where 
\begin{equation}
    (I_f)_a=\frac{\hbar}{4m_a}\int d{x}_adt\frac{\nabla_a\rho_a\cdot\nabla_a\rho_a}{\rho_a}; \text{   } (I_f)_b = \frac{\hbar}{4m_b}\int d{x}_bdt\frac{\nabla_b\rho_b\cdot\nabla_b\rho_b}{\rho_b}.
\end{equation}

\section{Proof of Theorem 1}
The starting point to prove Theorem is the Schr\"{o}dinger equation (\ref{jointSE}), which is derived based on the least observability principle and the two conditions mentioned in Theorem 1. Once obtained (\ref{jointSE}), proper mathematical tools in standard quantum mechanics can be applied. For instance, we can introduce a linear operator, the Hamiltonian operator $\hat{H} = -\frac{\hbar^2}{2m_a}\nabla^2_a-\frac{\hbar^2}{2m_b}\nabla^2_b + V$, and rewrite (\ref{jointSE}) as
\begin{equation}
    \label{SE2}
    \Psi(x_a, x_b, t)=e^{-i\hat{H}t/\hbar}\Psi(x_a, x_b, 0),
\end{equation}
where $t>t_0$ and we choose $t_0=0$.
\begin{lemma}
For a non-interaction bipartite system, $\Psi(x_a, x_b, t)$ is separable if and only if $\Psi(x_a, x_b, t_0)$ is separable.
\end{lemma}
Proof: Since $V(x_a, x_b)=V_a(x_a)+V_b(x_b)$, the Hamiltonian operator is separable $\hat{H}=\hat{H}_a+\hat{H}_b$, with $\hat{H}_a=-\frac{\hbar^2}{2m_a}\nabla^2_a+V_a$ and $\hat{H}_b=-\frac{\hbar^2}{2m_b}\nabla^2_b+V_b$. If $\Psi(x_a, x_b, 0)=\psi(x_a, 0)\phi(x_b,0)$, then from (\ref{SE2}),
\begin{equation}
    \label{SE3}
    \Psi(x_a, x_b, t)=e^{-i(\hat{H}_a+\hat{H}_b)t/\hbar}\psi(x_a, 0)\phi(x_b,0)=[e^{-i\hat{H}_at/\hbar}\psi(x_a, 0)][e^{-i\hat{H}_bt/\hbar}\phi(x_b, 0)]=\psi(x_a, t)\phi(x_b,t).
\end{equation}
On the other hand, left multiplying $e^{i\hat{H}t/\hbar}$ with both sides of (\ref{SE2}) one gets the time reversal of the Schr\"{o}dinger equation
\begin{equation}
    \label{RSE}
    \Psi(x_a, x_b, 0)=e^{i\hat{H}t/\hbar}\Psi(x_a, x_b, t).
\end{equation}
If $\Psi(x_a, x_b, t)=\psi(x_a, t)\phi(x_b,t)$, one has
\begin{equation}
    \label{SE3}
    \Psi(x_a, x_b, 0)=e^{i(\hat{H}_a+\hat{H}_b)t/\hbar}\psi(x_a, t)\phi(x_b,t)=[e^{i\hat{H}_at/\hbar}\psi(x_a, t)][e^{i\hat{H}_bt/\hbar}\phi(x_b, t)]=\psi(x_a, 0)\phi(x_b,0).
\end{equation}
Lemma 1 shows that for a non-interacting bipartite system, the separability of the wave function $\Psi(x_a, x_b, t)$ is completely determined by its initial condition $\Psi(x_a, x_b, 0)$.
\begin{lemma}
For a non-interaction bipartite system, $\Psi(x_a, x_b, t)$ is separable if and only if $\rho(x_a, x_b, t)$ is separable.
\end{lemma}
Proof: Since $\Psi({x}_a, {x}_b, t) = \sqrt{\rho({x}_a, {x}_b, t)}e^{iS(x_a, x_b, t)/\hbar}$, if $\rho({x}_a, {x}_b, t)$ is inseparable, obviously $\Psi({x}_a, {x}_b, t)$ is also inseparable. Now suppose $\rho({x}_a, {x}_b, t)=\rho_a(x_a,t)\rho_b(x_b,t)$, the continuity equation becomes
\begin{equation}
    \label{ContEq2}
    \rho_b[\frac{\partial\rho_a}{\partial t}+\frac{1}{m_a}\nabla_a(\rho_a\nabla_a S)] + \rho_a[\frac{\partial\rho_b}{\partial t}+\frac{1}{m_b}\nabla_b(\rho_b\nabla_b S)] = 0.
\end{equation}
Since $\rho_a$ and $\rho_b$ are completely independent and can be any arbitrary probability density distribution, (\ref{ContEq2}) implies
\begin{align}
    \label{ContEq3}
    \frac{\partial\rho_a}{\partial t}+\frac{1}{m_a}\nabla_a(\rho_a\nabla_a S) &=0, \\
    \frac{\partial\rho_b}{\partial t}+\frac{1}{m_b}\nabla_b(\rho_b\nabla_b S) &= 0.
\end{align}
Rewrite (\ref{ContEq3}) as $\partial\rho_a/\partial t = \nabla_a(\rho_a\nabla_a S)/m_a$. Since $\rho_a$ is a function of $(x_a, t)$, it demands $\nabla_a S$ must be independent of $x_b$. Similarly, $\nabla_b S$ must be independent of $x_a$. These require $S(x_a, x_b, t)$ must be separated into two terms $S(x_a, x_b, t)=S_a(x_a, t)+S_b(x_b, t)$. We need to show that $S_a$ and $S_b$, when pairing with $\rho_a$ and $\rho_b$ respectively, are valid solution to the Schr\"{o}dinger equation of individual subsystem. To see this, when $S=S_a+S_b$, the total observable information (\ref{totalI2}) in Appendix A becomes
\begin{equation*}
\begin{split}
    I &= \frac{2}{\hbar}\int d{x}_ad{x}_bdt\rho\{\frac{\partial S_a}{\partial t} + \frac{\partial S_b}{\partial t} + \frac{1}{2m_a}(\nabla_a S_a)^2
    + \frac{1}{2m_b}(\nabla_b S_b)^2 + V_a+V_b+\frac{\hbar^2}{8m_a}\frac{(\nabla_a\rho_a)^2}{\rho_a^2} + \frac{\hbar^2}{8m_b}\frac{(\nabla_b\rho_b)^2}{\rho_b^2}\} \\
    &= \frac{2}{\hbar}\int d{x}_adt\rho_a\{\frac{\partial S_a}{\partial t} + \frac{1}{2m_a}(\nabla_a S_a)^2
    + V_a+\frac{\hbar^2}{8m_a}\frac{(\nabla_a\rho_a)^2}{\rho_a^2}\} + \frac{2}{\hbar}\int d{x}_adt\rho_b\{\frac{\partial S_b}{\partial t} + \frac{1}{2m_b}(\nabla_b S_b)^2
    + V_b+\frac{\hbar^2}{8m_b}\frac{(\nabla_b\rho_b)^2}{\rho_b^2}\}.
\end{split}
\end{equation*}
Variation over $S_a$ and $\rho_a$ results in
\begin{align*}
    &\frac{\partial\rho_a}{\partial t}+\frac{1}{m_a}\nabla_a(\rho_a\nabla_a S) =0, \\
    &\frac{\partial S_a}{\partial t} + \frac{1}{2m_a}(\nabla_a S_a)^2 + V_a - \frac{\hbar^2}{2m_a}\frac{\nabla_a^2\sqrt{\rho_a}}{\sqrt{\rho_a}} = 0.
\end{align*}
Defining $\psi(x_a, t)=\sqrt{\rho_a}e^{iS_a/\hbar}$, the above two equations are combined into the Schr\"{o}dinger equation for $\psi(x_a, t)$. Similarly, variation over $S_b$ and $\rho_b$ results in the Schr\"{o}dinger equation for $\phi(x_b, t)=\sqrt{\rho_b}e^{iS_b/\hbar}$. This confirms that the wave function $\Psi({x}_a, {x}_b, t) = \psi(x_a, t)\phi(x_b, t)$, which is separable.

Now we can prove Theorem 1. Applying Lemma 2 at $t_0=0$, we know that for a non-interaction bipartite system, $\Psi(x_a, x_b, 0)$ is separable if and only if $\rho(x_a, x_b, 0)$ is separable. Then by Lemma 1, $\Psi(x_a, x_b, t)$ is separable if and only if $\rho(x_a, x_b, 0)$ is separable. It is equivalent to state that $\Psi(x_a, x_b, t)$ is inseparable if and only if $\rho(x_a, x_b, 0)$ is inseparable.

\end{document}